\documentclass[aps,reprint,superscriptaddress,floatfix,nofootinbib]{revtex4-1} 
\pdfoutput=1
\usepackage{graphicx}
\usepackage{amssymb,amsfonts,amsmath}
\usepackage{txfonts}
\usepackage{url}

\def \etal {{et~al.}}
\def \lett #1{(\textbf{#1})}
\def \Met  {Methods}

\def \rg   {r_\mathrm{g}}
\def \Rg   {R_\mathrm{g}}
\def \Rs   {R_*}
\def \pmfc {P_\mathrm{MFC}}
\def \frecip {f_\mathrm{reciprocal}}
\def \hone {h_1}
\def \htwo {h_2}

\begin{document}

\author{James P.~Bagrow}
\email{\raggedright To whom correspondence should be addressed. E-mail: james.bagrow@northwestern.edu}
\affiliation{Engineering Sciences and Applied Mathematics, Northwestern University, Evanston, IL, USA}
\affiliation{Center for Complex Network Research, Northeastern University, Boston, MA, USA}
\author{Yu-Ru Lin}
\affiliation{College of Computer and Information Science, Northeastern University, Boston, MA, USA}
\affiliation{Institute for Quantitative Social Science, Harvard University, Cambridge, MA USA}

\title{Mesoscopic structure and social aspects of human mobility}
\date{June 4, 2012}

\begin{abstract} 
    The individual movements of large numbers of people are important in many
    contexts, from urban planning to disease spreading. Datasets that capture
    human mobility are now available and many interesting features have been
    discovered, including the ultra-slow spatial growth of individual mobility.
    However, the detailed substructures and spatiotemporal flows of
    mobility---the sets and sequences of visited locations---have not been well
    studied. We show that individual mobility is dominated by small groups of
    frequently visited, dynamically close locations, forming primary
    ``habitats'' capturing typical daily activity, along with subsidiary
    habitats representing additional travel.  These habitats do not correspond
    to typical contexts such as home or work.  The temporal evolution of
    mobility within habitats, which constitutes most motion, is universal across
    habitats and exhibits scaling patterns both distinct from all previous
    observations and unpredicted by current models.  The delay to enter
    subsidiary habitats is a primary factor in the spatiotemporal growth of
    human travel. Interestingly, habitats correlate with non-mobility dynamics
    such as communication activity, implying that habitats may influence
    processes such as information spreading and revealing new connections
    between human mobility and social networks.  
\end{abstract}

\maketitle

\section*{Introduction}
Understanding human movement is essential for a range of
society-wide technological problems and policy issues, from urban
planning~\cite{horner2001embedding} and traffic
forecasting~\cite{kitamura2000micro}, to the modeling and simulation of
epidemics~\cite{pastor2001epidemic,hufnagel2004forecast,colizza2006role}.
Recent studies on mobility patterns have shown that spatiotemporal traces are highly
non-random~\cite{brockmann2006scaling,gonzalez2008understanding,song2010limits},
exhibiting distinct dynamics subject to geographic
constraints~\cite{eagle2009inferring,crandall2010inferring,WangKDD2011,CalabreseMobilePhonePLOSONE2011,expert2011uncovering,hui2005pocket}.
Analytical models have been developed to reflect individual mobility dynamics such as
the tendency to move back and forth between fixed locations on a regular
basis~\cite{song2010modelling}.
When examining populations, movement patterns may be highly correlated
with dynamics such as contact
preference~\cite{eagle2009inferring,WangKDD2011}, 
yet this has not been well studied at the individual level.
Previous work on human mobility has focused primarily on simple measures
that forego the majority of the detailed information available in existing data.
There is good reason for this, as basic approaches tend to be most fruitful for
new problems.  Yet these measures reduce an entire mobility pattern to a single
scalar quantity, potentially missing important details and throwing away crucial
information.

A number of approaches are available for studying the geographic substructure of
individual mobility.  One route is to perform spatial
clustering~\cite{jain1999dataClustering} on the specific locations an individual 
visits, potentially revealing important, related groups of locations.  However,
such analysis is purely spatial, neglecting the detailed spatiotemporal
trajectories available for each person, reducing their mobility to a collection of
geographic points and ignoring any information regarding the \textit{flows}, or
frequencies of movement, between particular locations.  At the same time, the
raw spatial distance separating two locations may not be meaningful: a short walk
and a short car trip typically cover very different distances in the same amount of time,
and the cognitive and economic costs associated with air travel depend only
mildly (if at all) upon distance~\cite{brons2002price}.  Modeling frameworks
such as the Theory of Intervening Opportunities~\cite{stouffer1940intervening}
and the recently introduced Radiation model~\cite{radiationModel} further argue
that raw distances are not necessarily the most effective determinant for
travel.
In this work we show the importance of incorporating how
frequently an individual travels between two locations, which naturally accounts for
spatial and dynamic effects while revealing the underlying spatiotemporal
features of human mobility.

\section*{Results}

\begin{figure*}[t!]
    \centerline{
        \includegraphics[width=\textwidth]{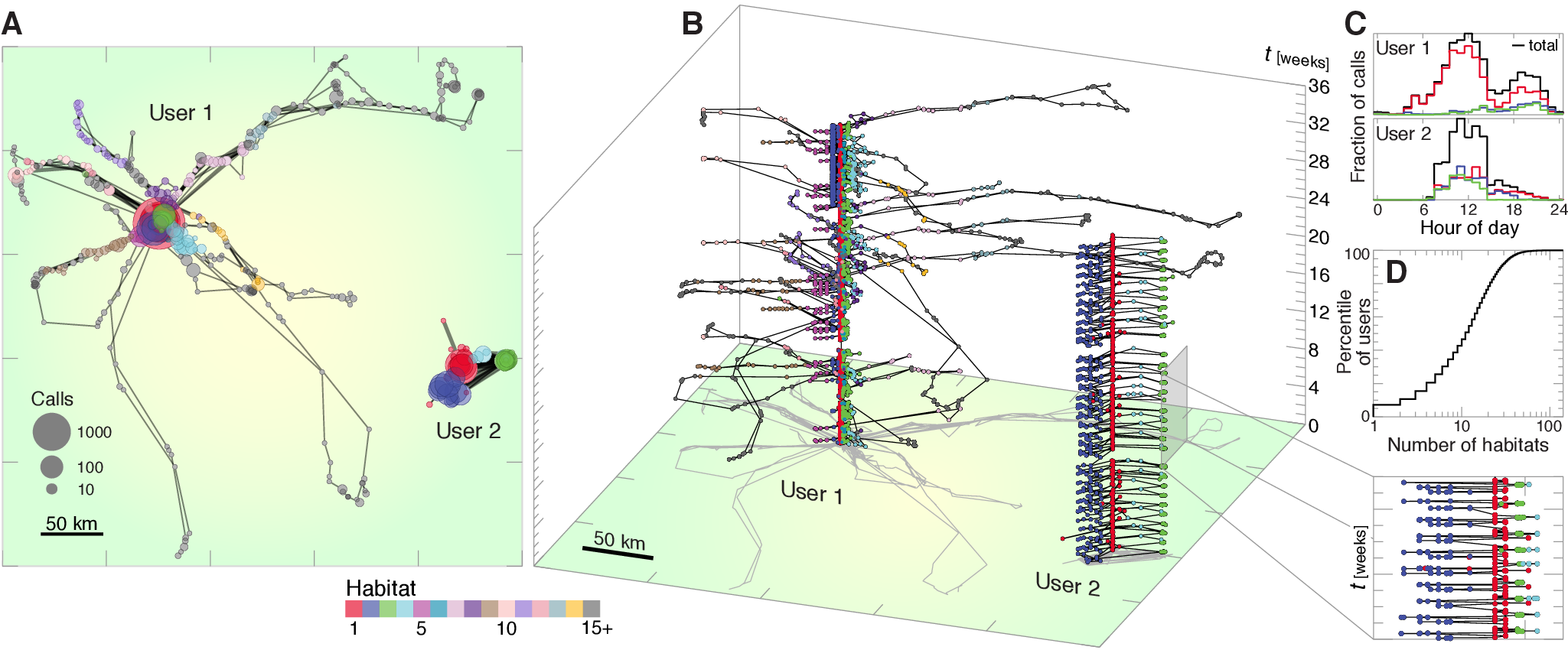}
    }
    \caption{\textbf{Habitats reveal the spatiotemporal substructure of human
        mobility patterns.}
        \lett{A} %
        Spatial trajectories of two users, one traveling to a large number of
        locations and another covering a smaller range.  Node size indicates the
        amount of time spent at a particular location (as quantified by mobile phone
        activity), node color represents the location's habitat detected using
        Infomap (see Methods), and line width approximates the number of trips
        between locations.  Habitats are ordered by call volume such that Habitat
        1 contains the most calls. 
        \lett{B} %
        Exploding the spatial trajectories from A in time (vertical axis), the
        recurrent nature of human mobility becomes evident, with a number of
        trips featuring both consistent destinations and consistently repetitive
        occurrence (zoom).  These features are the root cause of the high
        predictability that human motion is known to possess.
        \lett{C} %
        The daily call dynamics of the three most active habitats, as well as
        the overall dynamics (summed over all habitats).   The primary habitat
        contains the majority of temporal activity.  We see that User 1 tends to
        occupy his or her second and third habitats primarily at night, while
        User 2 is more evenly distributed.
        \lett{D} %
        The distribution of the number of habitats per user. The median number
        of habitats is 11.
        Due to their typical heterogeneity, we characterize population distributions
        using percentiles, proportional to the cumulative distribution.
    }
    \label{fig:tree}
\end{figure*}

Beginning from a country-wide mobile phone
dataset~\cite{onnela2007structure,gonzalez2008understanding,bagrow2009investigating,song2010limits,song2010modelling,park2010eigenmode,bagrowDisaster2011pone,10.1371/journal.pone.0016939},
we extract 34 weeks of call activity for a sample population of approximately 90
thousand phone users.  Each call activity time series encodes the spatiotemporal
trajectory of that user.  (See Materials and Methods and Supporting
Information (File S1) for details about the data.)  For each user we construct a
directed, weighted \textit{mobility network} capturing the detailed flows
between individual locations (represented using cellular towers).  Examples of
both mobility networks and spatiotemporal mobility flows are shown in
Figs.~\ref{fig:tree}A and B, respectively. The recurrent and repetitive nature
of human motion is clearly visible in Fig.~\ref{fig:tree}B, where we explode the
user trajectories vertically in time.
We apply to each user's mobility network an information-theoretic graph
partitioning method known as Infomap~\cite{rosvall2008maps}, which uses the
flows of random walkers to find groups of dynamically related nodes in directed,
weighted networks.  We do not use spatial or distance information in
partitioning, instead Infomap mirrors the stochastic process underlying human
mobility flows; see File S1 Sec.~S3 for details.  (Infomap's underlying mechanism is
further justified in this context by the results of \cite{park2010eigenmode}.)
The groups of locations that we discover, which we refer to as mobility
``habitats,'' will be shown to be crucial to both the spatiotemporal dynamics of
human motion, and to the interplay between mobility and human interaction
patterns.  
We rank habitats in decreasing order of phone activity, such that a user's most
frequently visited habitat is Habitat 1 or the primary habitat.  We observe that
human mobility is almost universally dominated by the primary habitat, where the
majority of user call activity occurs---and thus it incorporates both home
and work, home and school, or other major social contexts---along with a number of less active
subsidiary habitats (see Fig.~\ref{fig:tree}C, File S1 Fig.~B, Sec.~S3.2).
We further see in Fig.~\ref{fig:tree}D that most users possess 5--20 habitats,
while only approximately 7\% of users have a single habitat.
Note that these habitats, unique for each member of the population, differ
greatly from existing work on partitioning mobility or social
connectivity~\cite{brockmannEffectiveBordersPLOSONE2011,expert2011uncovering,ratti2011Britainplosone},
which instead focus entirely on partitioning a single geographic network
aggregated from large populations.

\subsection*{Spatial characteristics}

\begin{figure*}[t]
    \centerline{\includegraphics[]{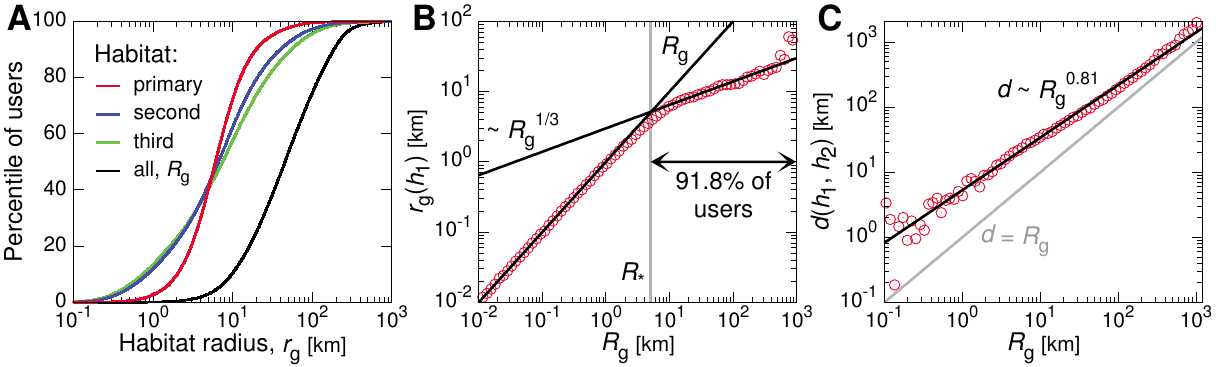}}
    \caption{ \textbf{Spatial properties of mobility habitats.} %
        We characterize each habitat's spatial extent by computing the radius of
        gyration $\rg(h)$ considering only calls placed from locations within habitat $h$.
        \lett{A} %
        The distribution of habitat radii over the population shows that the
        primary habitat tends to be more spatially compact than the less
        frequented habitats, though most are consistently smaller than the total
        $\Rg$ computed using all phone activity.  
        \lett{B} %
        The growth in the radius of the primary habitat $\rg(\hone)$ as a
        function of total radius $\Rg$.  For $\Rg < \Rs \approx 5$ km, we see
        $\rg(\hone) \approx \Rg$, indicating that those users are characterized
        by a single habitat.  In contrast, $\rg(\hone) \sim \Rg^{1/3}$ for $\Rg
        > \Rs$. Since approximately 92\% of the population have $\Rg > 5 $ km,
        the majority of users exist in a regime where their primary habitat
        encompasses a potentially far smaller spatial region than their total
        mobility. %
        \lett{C} %
        For users with multiple habitats, the distance $d(\hone,\htwo)$ between
        the first and second habitat's centers of mass is consistently greater
        than $\Rg$ (grey line) and exhibits power law scaling, $d(\hone,\htwo)
        \sim \Rg^\beta$, with $\beta = 0.81 \pm 0.02$.  
        Taken together, we see that most habitats are both well separated and
        spatially compact, and that the magnitude of $\Rg$ is primarily due to
        movement between these habitats. 
    \label{fig:spatial}}
\end{figure*}

The spatial extent of a user's total mobility pattern has been shown to be well
summarized by a single scalar quantity, the radius of gyration, or gyradius,
$\Rg^2 = \left< \left|\mathbf{r}_i - \mathbf{r}_\mathrm{CM}\right|^2\right>_i$,
where $\mathbf{r}_i$ is the spatial position of phone call $i$ and
$\mathbf{r}_\mathrm{CM}$ is the user's center of mass
\cite{gonzalez2008understanding}. In addition to using the global gyradius we
also compute the reduced radius of gyration $\rg(h)$ for each habitat $h$,
considering only those locations and calls contained within that habitat.  In
Fig.~\ref{fig:spatial}A we plot the population distributions of the first three
habitat's $\rg$, compared with the total gyradius $\Rg$ considering all calls
placed from all visited locations.  
This shows that the spatial extent of habitats tends to be far smaller than the
total mobility, often by an order of magnitude, and that most users have a
habitat $\rg$ between $1$--$10$ km. See also File S1 Fig.~D.
In Fig.~\ref{fig:spatial}B we study the functional dependence of the primary
habitat's gyradius, $\rg(\hone)$, versus $\Rg$. We uncover an intriguing power
law scaling relation characterized by two regimes, where $\rg(\hone) \sim
\Rg^\alpha$ with $\alpha=1$ for $\Rg < \Rs \approx 5$ km, and $\alpha=1/3$ for
$\Rg>\Rs$.  The linear relationship below this critical radius $\Rs$
indicates that those users (roughly 8\% of the population) are
mostly characterized by a single habitat.
(In fact, only 54.8\% of users with $\Rg < 5$ km have one habitat,
but that 97.6\% of their calls on average occur within their primary habitat.)
But once a user's range extends beyond this critical 5 km cutoff (true for 92\%
of the population) a new regime emerges where multiple habitats exist and tend
to be far smaller and more spatially cohesive than the total mobility (since
$\alpha < 1$).
(For users with $\Rg > 5$ km, only 2.9\% have one habitat and the primary
habitat accounts for 78.7\% of activity on average.)
Finally, in Fig.~\ref{fig:spatial}C we show the geographic distance
$d(\hone,\htwo)$ between the centers of mass of the two most heavily occupied
habitats, as a function of $\Rg$.  This also exhibits a power law scaling,
$d(\hone,\htwo) \sim \Rg^\beta$ with $\beta = 0.81 \pm 0.02$.  These distances
tend to be far larger than the total $\Rg$ (gray line), indicating that the
magnitude of $\Rg$ is primarily determined by movement between spatially
cohesive and well separated habitats.

\subsection*{Temporal characteristics}

Given the importance of habitats to the spatial extent of human motion, one must
ask: how do these habitats form and evolve over time? To what extent are the
temporal dynamics of human movement reflected in the evolution of these
habitats?  Recently, considerable effort has been undertaken to understand the
intriguing temporal features of human mobility, including the previously
observed ultra-slow growth in time $t$ of $\Rg \sim \left(\log t\right)^\gamma$,
with $\gamma > 1$~\cite{gonzalez2008understanding,song2010modelling}.  Given the
contribution of habitats to the magnitude of $\Rg$, shown in
Fig.~\ref{fig:spatial}, a primary question becomes: how do habitats impact these
temporal features?  For example, how do individual habitat $\rg$'s evolve over
time, compared with that of the total $\Rg$? 

\begin{figure}[t!]
    \centerline{\includegraphics[]{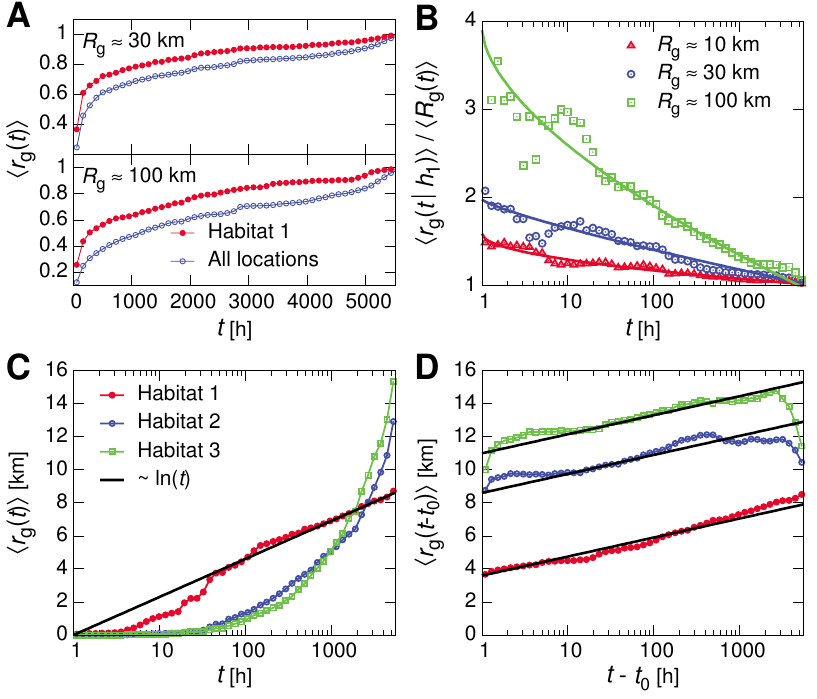}}
    \caption{ \textbf{Temporal evolution of human mobility.} %
        \lett{A} %
        The time evolution of $\rg(h_1)$ compared with $\Rg$, where both are
        normalized by their final values at the end of the observation window.
        We see that the primary habitat tends to reach saturation faster than
        the overall gyradius, indicating different temporal dynamics. %
        \lett{B} %
        To quantify the saturation rate, we plot the ratio of the two curves
        from A, for groups of users with different $\Rg$.  We see that the
        primary habitat saturates more quickly as the overall $\Rg$ grows. Solid
        lines of the form $\sim (\log t)^\epsilon$ provide a guide for the eye. %
        \lett{C} %
        The unnormalized growth in habitat size for the first three habitats.
        The primary habitat shows a distinct, approximately logarithmic temporal
        scaling. The other habitats show a longer delay before $\rg$ begins to
        grow polylogarithmically. %
        \lett{D} %
        Given this delay, we now shift the time series of $\rg(h)$ for each habitat by
        $t_0(h)$, the time when the user first entered habitat $h$.  Doing so we
        recover pure logarithmic scaling for all habitats, $\rg \sim \log
        \left(t-t_0\right)$,
        indicating that a major factor in the scaling of human mobility is the
        delay it takes for a user to transition to his or her non-primary
        habitats. %
   \label{fig:temporal_rg}}
\end{figure}

In Fig.~\ref{fig:temporal_rg} we study the temporal evolution of $\rg$ and $\Rg$
by considering only those calls occurring up to time $t$, from either individual
habitats or all locations, where $t=0$ is the time of the user's first call. In
Fig.~\ref{fig:temporal_rg}A we plot the time series of
$\rg(\hone)$ and $\Rg$, normalized by the final values of each respective
series. We observe that $\rg$ saturates at its final value more quickly than the total
mobility's $\Rg$.  To further quantify this saturation, we plot in
Fig.~\ref{fig:temporal_rg}B the ratio between $\rg(\hone)$ and $\Rg$ as a
function of time, for groups of users with different final values of $\Rg$.  We
observe increasingly rapid saturation of $\rg$ as the total $\Rg$ increases.
This implies that the primary habitat is explored more quickly than the total
extent of a user's mobility pattern and that users who cover large distances
explore their primary habitat more quickly relative to their total mobility than
users who traverse relatively smaller regions. This is particularly interesting
as one may initially expect such exploration to be at a constant rate relative
to their total $\Rg$.
In Fig.~\ref{fig:temporal_rg}C we study the temporal evolution of $\rg(t)$ for the
first three habitats, averaged over users with $\Rg \approx 30$ km.  We
observe approximately logarithmic growth, $\rg(\hone) \sim \log t$, for the primary
habitat (slower growth than that observed in
\cite{gonzalez2008understanding,song2010modelling}) while subsidiary habitats'
gyradii $\sim (\log t)^\delta$, with $\delta > 1$ (growth more similar to
\cite{gonzalez2008understanding,song2010modelling}).  However, this analysis
neglects an important detail: users do not begin exploring all of their habitats at
the same time. Therefore 
in Fig.~\ref{fig:temporal_rg}D we plot the same population-averaged radii, but
we now individually shift each user's time series of $\rg(h)$ by a time
$t_0(h)$, the time the user first entered habitat $h$, not simply made his or
her first global call.  Doing so accounts for the waiting times for users to
visit habitats within our observation window.  With this crucial correction we
reveal \emph{for all habitats} purely logarithmic growth in $\rg$, implying a
universality in the exploration of habitats (which differ only in their overall
spatial scale, with the primary habitat tending to be the most compact).  Thus,
the polylogarithmic growth of $\Rg$, where $\Rg$ is initially small then grows
faster than logarithmic in time, is primarily due to the temporal delay
it takes users to exit their respective primary habitats and then rapidly
traverse a relatively large distance to reach their other habitats.
We further study these habitat entrance times in File S1, Sec.~S3.2 and Fig.~E.

\subsection*{Social characteristics}

\begin{figure}[t!]
    \centerline{\includegraphics[]{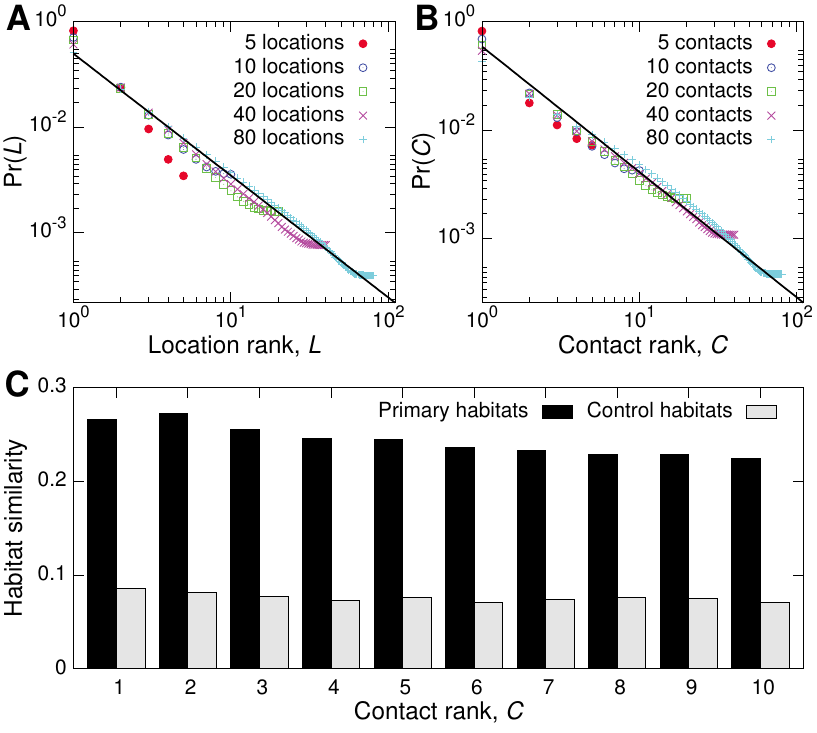}} %
    \caption{ \textbf{Contact activity and habitats.}  
        \lett{A} %
        The Zipf law governing the probability $\mathrm{Pr}(L)$ for a user to
        visit his or her $L$-th most visited location, as first observed by
        {\protect Gonz\'alez} \etal{}~\cite{gonzalez2008understanding}. The
        solid line indicates $\mathrm{Pr}(L) \sim L^{-1.5}$ 
        \lett{B} %
        Interestingly, we observe an identical Zipf law for the probability
        $\mathrm{Pr}(C)\sim C^{-1.5}$ for a user to call his or her $C$-th most
        contacted partner.  This holds regardless of the total number of
        contacts for a user.  This implies that the same underlying mechanism
        may govern how users choose both locations to visit and friends to
        contact.
        \lett{C} %
        The habitat similarity, related to the number of common locations,
        between a user's primary habitat and the primary habitat of their
        contacts, averaged over pairs where both users are present in our data.
        We see that, despite the Zipf law in B, users' habitats tend to be
        surprisingly similar to their most contacted ties, even for those less
        frequently contacted users.  Control habitats, generated by randomly
        shuffling a user's visited locations between his or her original
        habitats, exhibit lower similarity.  See \Met{} for habitat similarity
        and controls.
    \label{fig:zipf_habsim}}
\end{figure}

Finally, a major question in the realm of mobility and human dynamics is the
connection between spatiotemporal dynamics and other activity
patterns~\cite{eagle2009inferring}.  For example, information spreading in
heterogeneous systems of agents is a process that involves both the
spatiotemporal mobility of the agents and their long-range communication
activities.
In this context, would the currently occupied habitat affect or be affected by how a
user chooses a particular communication partner to engage?  Such questions can
also be addressed with mobile phone data, where phone communications capture a
primary mode of information diffusion on the underlying social
network~\cite{onnela2007structure}. To begin, we first recall a
result from Gonz\'alez \etal{}~\cite{gonzalez2008understanding}.  They found that
users occupy locations following a Zipf law, where the probability
$\mathrm{Pr}(L)$ to visit the $L$-th most frequented location follows
$\mathrm{Pr}(L) \sim L^{-1.5}$.  We reproduce this result in
Fig.~\ref{fig:zipf_habsim}A.  Interestingly, we discover
(Fig.~\ref{fig:zipf_habsim}B) a potentially identical mechanism for how users
choose to contact their communication partners, i.e.~the probability $\mathrm{Pr}(C)$ for a
user to call his or her $C$-th most contacted partner also follows
$\mathrm{Pr}(C) \sim C^{-1.5}$.  See also~\cite{backstrom2011center}.  Finally,
a number of users within our population have contacts that are also
within the population, meaning we have habitat information for both users.  An
interesting question is: how similar are the habitats of users in close
communication, and will this similarity be lower for pairs with less frequent
interaction?  We measure the similarity between the primary habitats of pairs
of users interacting with one another by computing the relative number of
locations the habitats have in common (see Methods and materials).  In
Fig.~\ref{fig:zipf_habsim}C we plot this Habitat similarity as a function of
contact rank $C$. 
We see that, despite the Zipf law in Fig.~\ref{fig:zipf_habsim}B, users' primary
habitats tend to be highly similar to the primary habitats of their most
contacted ties. Nevertheless, there is little dependence on contact
rank: one partner that is contacted an order of magnitude less often than another
has almost the same primary habitat similarity.  In other words, it takes
very little communication to generate considerable habitat
overlap~\cite{crandall2010inferring}.  Meanwhile, control habitats, generated by
randomly distributing each user's visited locations between their habitats (see
Methods and materials), show smaller similarity and no effective trend.

\begin{figure}
    \centerline{\includegraphics[]{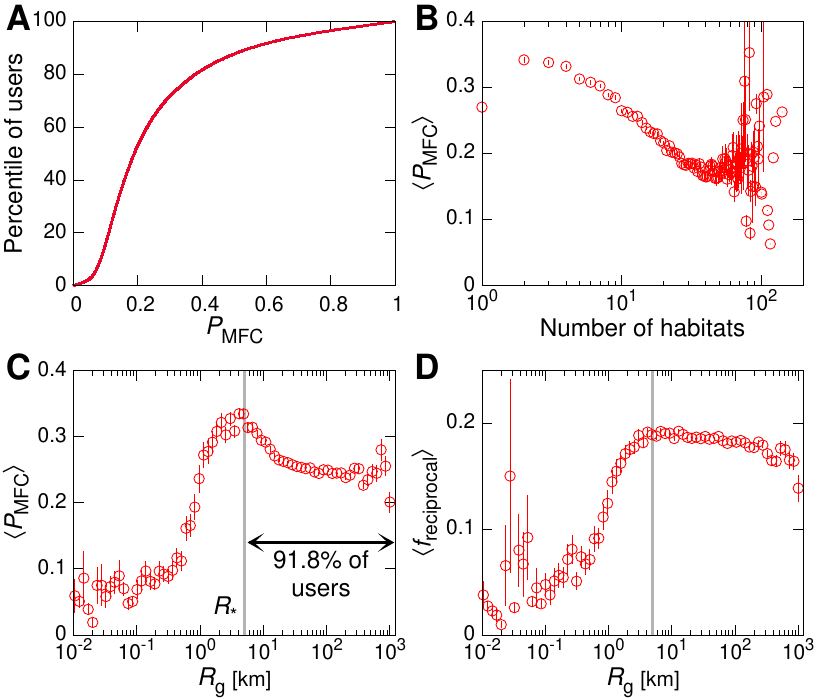}} %
    \caption{\textbf{Communication and mobility dynamics.} %
        We characterize the interaction concentration of a user by
        $\pmfc$, the probability for that user to place a call to his
        or her Most Frequent Contact. 
        \lett{A} %
        The distribution of $\pmfc$ over the population shows that most users
        have $\pmfc$ between approximately $0.1$ and $0.4$.  (See also
        Fig.~\ref{fig:zipf_habsim}b.)
        \lett{B} %
        To connect the concentration with user mobility, we study how the mean
        $\pmfc$ varies with the number of habitats each user possesses.  We see
        that $\pmfc$ gradually decays as the number of habitats grows,
        indicating that broadly traveled individuals tend to more evenly
        distribute their calls over their partners.
        \lett{C} %
        Studying $\pmfc$ as a function of $\Rg$, we uncover an intriguing
        relationship.  For users with particularly small mobility 
        ranges, 
        $\pmfc$ is small but grows as $\Rg$ grows.  This continues until
        $\Rg\approx\Rs$, the same critical radius size observed in
        Fig.~\ref{fig:spatial}b.  The mean $\pmfc$ then decays for $\Rg > \Rs$.
        Surprisingly, this implies that the distribution of call activity over a
        user's partners exhibits different behavior depending on whether that
        user possess one mobility habitat, or many habitats.
        \lett{D} %
        The fraction of reciprocated contacts $\frecip$ as a function of $\Rg$
        shows a trend similar to $\pmfc$. Not only do those users
        with small $\Rg$ tend to be distinctly less socially concentrated
        compared with most users, they also tend to make more non-reciprocated
        contacts (see File S1 Fig.~C for details).
        Error bars indicate $\pm 1$ s.e.
    \label{fig:plot_PMFC}}
\end{figure}

We further characterize the ``interaction concentration'' of a user by introducing
$\pmfc$, the probability that the next call placed by the user goes to that
user's Most Frequent Contact, the partner that is most often in contact with the
user.
Users with a small $\pmfc$ tend to distribute their calling activity more evenly
across their partners, while users with large $\pmfc$ are more concentrated
and focus much of their attention upon a single individual. 
In Fig.~\ref{fig:plot_PMFC} we study how $\pmfc$ depends on the properties of a
user's mobility pattern. First, in Fig.~\ref{fig:plot_PMFC}A we show the
distribution of $\pmfc$ over the user population.  Most users possess $0.1 \leq
\pmfc \leq 0.4$ while very few users have either very small or very large $\pmfc$.  
In Fig.~\ref{fig:plot_PMFC}B we connect this interaction concentration with the
user mobility patterns by showing that the mean $\pmfc$ decays as the number of
habitats a user visits grows.  This means that users who travel broadly, leading
to complex mobility patterns and multiple habitats, tend to distribute their
communication activity more uniformly over their contacts.  
Next, in Fig.~\ref{fig:plot_PMFC}C we quantify how $\pmfc$ varies with the
total gyradius $\Rg$.  We see an intriguing connection to a previous result:
For users with small $\Rg$, the $\pmfc$ is small but grows as $\Rg$ grows.  This
continues until $\Rg \approx \Rs$, the same critical radius that appeared in
Fig.~\ref{fig:spatial}B.  Above $\Rs$, we see that $\pmfc$ now slowly decays
with $\Rg$.
To further understand this, we plot in Fig.~\ref{fig:plot_PMFC}D the fraction of
reciprocated contacts $\frecip$ (see Materials and methods) as a function of
$\Rg$.  The plot exhibits a roughly similar trend as Fig.~\ref{fig:plot_PMFC}C:
$\frecip$ grows while $\Rg < \Rs$ then, above the same critical radius,
$\frecip$ decays slowly with $\Rg$. This decay relative to the peak value at
$\Rg \approx \Rs$ is slower for $\frecip$ than for $\pmfc$. 

Taken together, Figs.~\ref{fig:plot_PMFC}C and D show that when $\Rg > \Rs$,
user communication activity---both how much they concentrate upon their MFCs and how
many of their ties are reciprocated---depends only weakly on $\Rg$.  However,
those users with low $\Rg$ tend to show distinctly different behavior, both
being less concentrated on their MFCs compared with most users, and making a
larger number of non-reciprocated contacts (File S1 Fig.~C).
Since users with $\Rg < \Rs$ primarily possess a single habitat, these results
imply that the mechanism governing how users distribute their activity over
their contacted partners may differ for those users with a
single habitat compared with those users whose mobility leads to multiple
habitats.  
We used Kendall rank correlation and associated hypothesis
tests~\cite{kendall90rank} to verify the statistical validity of the observed
relationships. 
See File S1 Sec.~S6 and Table A for hypothesis tests between these and additional
measures.

The mobile phone data also provides demographic information for the majority of
users, specifically self-reported age and gender. In File S1 Sec.~S4 we study the
results of Fig.~\ref{fig:plot_PMFC} after decomposing the sample into age and
gender groups.   One may expect these results to change when focusing on these
different groups. Yet in Figs.~F and G we find qualitatively similar results
to Fig.~\ref{fig:plot_PMFC} with only small differences: $\pmfc$ tends to be
slightly higher for women than for men, and increases with user age.  After
considering these demographic dependencies on $\pmfc$, we observe the same
relationships between communication activity and mobility.

\section*{Discussion}

We have shown that accurately understanding human mobility requires an analysis
using the complete spatiotemporal flows captured for each user. Basic measures
such as the gyradius $\Rg$ constitute an excellent starting point, but such
single scalar quantities simply cannot capture the full complexity of an
individual mobility pattern. As the quality and quantity of available data
increases, we expect our understanding of the various factors shaping human
mobility to continue to improve.

Given that users spend the majority of time occupying their primary habitats,
understanding the detailed features of the primary habitat will be crucial for
applications such as search and rescue during
emergencies~\cite{bagrowDisaster2011pone} or containing the spread of epidemic
diseases~\cite{pastor2001epidemic,hufnagel2004forecast,colizza2006role}, since
most users will be within their primary habitats at the onset of such events.
Meanwhile, detailed information regarding unusual trips away from the primary
habitat may prove useful both for curtailing diseases and for optimizing
transportation infrastructure and energy usage.  Likewise, the universal
logarithmic scaling laws for intra-habitat mobility uncovered in
Fig.~\ref{fig:temporal_rg}D are not accounted for by current modeling
frameworks~\cite{song2010modelling}; more effort may be necessary to acceptably
model the microscopic structure of individual human motion.  
The connections we reveal between communication dynamics and human mobility may have
important consequences for understanding the spread of information or rumors
through a population, as such processes may spread both spatially and
socially~\cite{kleinberg2000navigation}.
Further investigation of such connections may prove fruitful in a number of
areas, including information diffusion and social contagion.

\section*{Materials and Methods}

\subsection*{Dataset}  

We use a large-scale, de-identified mobile phone dataset, previously studied
in~\cite{onnela2007structure,gonzalez2008understanding,bagrow2009investigating,song2010limits,song2010modelling,park2010eigenmode,bagrowDisaster2011pone}.
We sample approximately $90$ thousand users from the total dataset, according to
the activity criteria introduced in~\cite{song2010limits} (see also File S1 Fig.~H).
We retain nine months of phone activity for each user. A ``call'' is either a
text message or a voice call, and we use the cellular tower that handled the
communication to represent the location $L(t)$ of a call made at time $t$. Call
times are kept at an hourly resolution. The coordinates of these towers are used
to compute the radii of gyration~\cite{gonzalez2008understanding}. Phone call
recipients determine the communication partners of a user.  Since a single phone
call between two individuals may not represent a meaningful tie, we consider
user $B$ to be a partner of user $A$ only if we observe at least one
reciprocated pair of calls ($A$ called $B$ and $B$ called
$A$)~\cite{onnela2007structure}.  We do not require user $B$ to be in our sample
population, except when we compute habitat similarity.
We define the fraction of reciprocal ties for user $A$ as $\frecip(A) = \sum_B
X(A,B) X(B,A) / \sum_B X(A,B)$ where $X(A,B)=1$ if $A$ contacted $B$ at least
once, and zero otherwise.

\subsection*{Finding mobility habitats}

For each user we convert their trajectory $\xi = \{L(t_1), L(t_2), ...\}$, with
$t_i > t_{i'}$ for $i > i'$, into a weighted digraph where the weight on link
$L_i \to L_j$ represents the number of times the ordered pair of locations
$(L_i,L_j)$ was observed in $\xi$ (File S1 Fig.~A).  The community discovery
method Infomap \cite{rosvall2008maps} was applied to each digraph, using the
default parameters (10 attempts and self-loops ignored). The discovered groups
of locations are the habitats for that user.  Habitats are ranked by total
number of calls.

\subsection*{Habitat similarity}

For a user $A$ with contact $B$, both present in our sample, we define
the similarity $S(A,B)$ between their primary habitats to be the Jaccard
coefficient between the sets of locations comprising those habitats.  If these
sets are disjoint $S(A,B) = 0$, whereas $S(A,B) = 1$ if they overlap completely.

\subsection*{Controls}

It is important to understand the significance of the results we have presented
here, in particular whether the results associated with the habitats we find are
meaningful.  We compute null or control habitats, generated for each user, by
randomly assigning that user's visited locations to habitats while preserving
the number of habitats and the number of locations within each habitat.  This
strictly controls for the habitat size distributions while testing the effects
of habitat membership.  In File S1 Fig.~I we further show that the pure
logarithmic time evolution is absent in control habitats, indicating that the
temporal evolution we observe in Fig.~\ref{fig:temporal_rg}D is not due to the
relative sizes (numbers of locations) of the habitats, nor to simply the number
of habitats, but due more fundamentally to their spatial structure and the
spatiotemporal flows of the users.  In Fig.~\ref{fig:zipf_habsim} we see that
these control habitats have lower similarity than the actual habitats.  See File
S1 Sec.~S6 for details.

\section*{Acknowledgments}

We thank F.~Simini, J.~Menche, J.-P.~Onnela, S.~Lehmann, W.-T.~Chung,
D.~Brockmann, B.~Uzzi, and D.~Lazer for many useful discussions and
A.-L.~Barab\'asi, D.~Lazer, B.~Uzzi and D.~Brockmann for support.  We gratefully
acknowledge the hospitality of the LazerLab at Northeastern University,
supported by MURI grant \#G00003585. Any opinions, findings, and conclusions or
recommendations expressed in this material are those of the authors and do not
necessarily reflect the views of the funding source.

\end{document}


\renewcommand{\thefigure}{\Alph{figure}}
\renewcommand{\thesection}{S\arabic{section}}
\renewcommand{\thetable}{\Alph{table}}
\renewcommand{\theequation}{S\arabic{equation}}
\renewcommand{\thesubfigure}{\alph{subfigure}}

\noindent{\huge Supporting Information}
\baselineskip24pt
\singlespacing
\noindent\textit{Mesoscopic structure and social aspects of human mobility} \\
\noindent by James P.~Bagrow and Yu-Ru Lin

{\small
\renewcommand*\contentsname{Table of Contents}
\tableofcontents
\listoffigures
\listoftables
}

\section{Dataset}\label{sec:dataset}
We use a set of de-identified billing records from a Western European mobile
phone service
provider~\cite{onnela2007structure,gonzalez2008understanding,song2010limits,song2010modelling,bagrow2009investigating,bagrowDisaster2011pone}.
The records cover approximately 10M subscribers within a single country over 3
years of activity.  Each billing record, for voice and text services, contains
the unique identifiers of the caller placing the call and the callee receiving
the call; an identifier for the cellular antenna (tower) that handled the call;
and the date and time when the call was placed.  Coupled with a dataset
describing the locations (latitude and longitude) of cellular towers, we have
the approximate location of the caller when placing the call. For this work we
do not distinguish between voice calls and text messages, and refer to either
communication type as a ``call.''

These phone records cover approximately 20\% of the country's mobile phone
market.  However, we also possess identification numbers for phones that are
outside the service provider but that make or receive calls to users within the
company.  While we do not possess any other information about these lines, nor
anything about their users or calls that are made to other numbers outside the
service provider, we do have records pertaining to all calls placed to or from
those ID numbers involving subscribers covered by our dataset. Thus egocentric
networks~\cite{Wasserman1994} between users within the company and their
immediate neighbors only are complete. This information was used to generate
egocentric communication networks and to study the MFC probability and its relationship
to human mobility patterns.

We generate a sample population of approximately 90k users (specifically,
$N=88137$), using the criteria introduced in \cite{song2010limits}.  Each user's
call history during our nine-month tracking period yields three time series: 
(i) event times for when calls are made, kept to an hourly resolution; 
(ii) locations of calls, as quantified by the cellular tower transmitting the
call; and 
(iii) communication partners who receive calls.  These time series allow us to
reconstruct both geographic trajectories and egocentric communication networks for each
user.

\section{Mobility networks}\label{sec:mobnets}

We construct for each sample user a weighted, directed mobility network $G$ (or
\textit{MobNet} for short) using the user's time-ordered trajectory $\{L(t_1),
    L(t_2), \ldots\}$, where $L(t)$ is the location the user called from at time
$t$.  Each link $(L_i \to L_j)$ in $G$ represents the user placing a call at
location $L_i$ followed by a call at location $L_j$.  The weight on link $(L_i
\to L_j)$ gives the number of times the user made that particular relocation
during the sample window.

\begin{figure}[t!]
  \centering
  \subfloat[]{\label{fig:gull}{\includegraphics[width=0.25\textwidth,trim=0 -30 0 30]{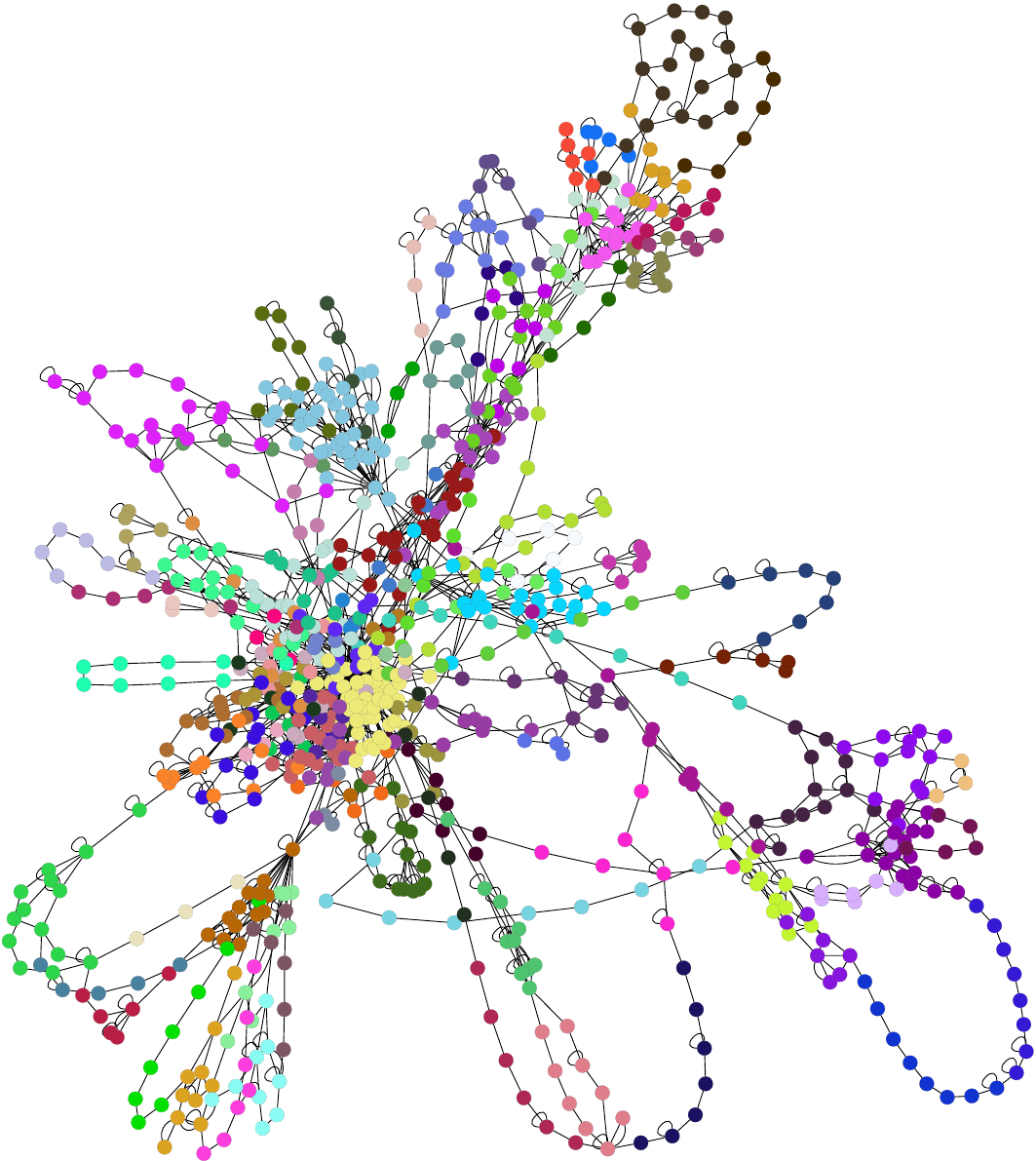}}}%
  \hspace{2em}
  \subfloat[]{\label{fig:mouse}\includegraphics[width=0.6\textwidth,trim=0 -11 0 11]{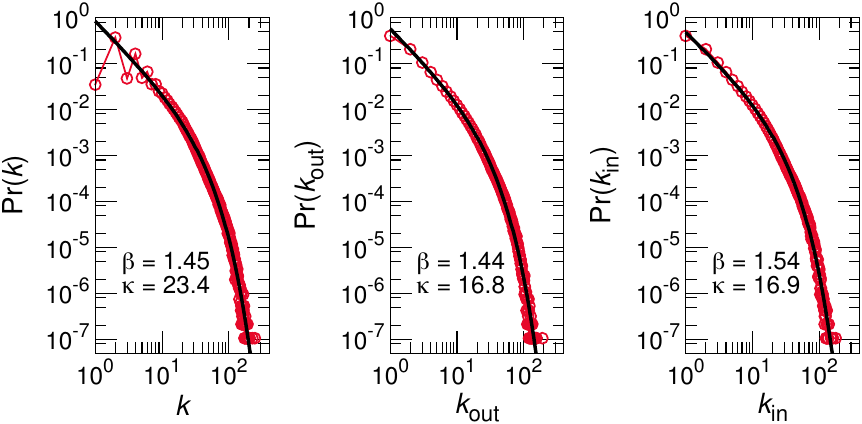}}\\
    \vspace{-1.5em}
        \caption[Properties of mobility networks.]{%
            \textbf{Properties of mobility networks.}
            \textbf{\subref{fig:gull}}
            An example mobility network (MobNet), drawn without using spatial
            coordinates.  Several dense cores are visible, as are a number of
            unusually long loops, representing one-time trips. Node colors
            indicate habitats, discovered using Infomap~\cite{rosvall2008maps}.
            Link weights and directions have been omitted for clarity.
            \textbf{\subref{fig:mouse}}
            Degree distributions (undirected, outdegree, and indegree) for all
            MobNets.  All are well described by a power law with an exponential
            cutoff, $\Pr(k) \sim k^{-\beta} e^{-k/\kappa}$, meaning that users
            typically visit many locations only a small number of times while a
            few locations are visited many times.  Note that here
            ``degree'' refers to the number of connections per
            location (the number of unique locations visited before or
            after visiting that location) not the number of
            communication partners a user has.
            \label{fig:example_mobnet}}
\end{figure}

In Fig.~\ref{fig:example_mobnet}a we draw the mobility network of a single user.
This network was drawn using a typical force-directed graph layout
algorithm~\cite{di1999graph} and does not use geographic tower locations.  We
see several dense cores comprising groups of frequently visited locations as
well as a number of long loops or chains representing sequential calls placed
during one-time, typically long-range trips.  These mobility networks feature
broad degree distributions, well described by a truncated power law of the form
$\Pr(k) \sim k^{-\beta} e^{-k/\kappa}$, for constants $\beta$ and $\kappa$,
where $k$ is the number of connections of a location (or number of unique
locations visited before or after visiting that location).  The nature of this
broad distribution is not surprising given the Zipf law for location selections,
observed in~\cite{gonzalez2008understanding} (see also Main text Fig.~4a).
Although one may not expect this distribution to hold for both in- and
out-degrees, we do observe similar patterns in our directed networks. 

\section{Mobility habitats}\label{sec:habitats}

In this work we start by identifying groups of related locations, for each user.
These groups may correspond to home, work, school, or any number of other
contexts throughout daily life. (In practice, we find that the most occupied
habitat accounts for the majority of activity and thus must be both home and
work, home and school, etc.)
The mobility networks we study are inherently spatial, possessing a unique
geographical embedding, yet we do not discover these groups through spatial
clustering methods, so it may be misleading to refer to these groups as
clusters.  Likewise, although we use a community detection method known as
Infomap~\cite{rosvall2008maps} to find these groups, mobility networks are not
social networks, so referring to these groups as communities may also be
misleading.  To avoid confusion, we instead term these groups ``habitats.''  We
rank each user's habitats by the total number of phone calls that occur within
the habitat, so that Habitat 1 is most active, Habitat 2 is second most active,
etc.  Habitats are not ranked by number of locations, though these tend to
correlate.  See Methods and materials in the main text for details on how to
apply Infomap.

\subsection{Justification for Infomap}\label{subsec:justifyInfomap}

There are numerous algorithms for detecting communities in networks
\cite{fortunato2010community}.  We believe Infomap to be ideally suited for our
purposes here, for two main reasons.
The first reason is that it is specifically capable of handling both weighted
and directed networks, and much of the focus of the original publication
\cite{rosvall2008maps}
was devoted to the sometimes confusing effects of community structure in
directed networks.  Infomap's theoretical basis rests upon the notion of random
walkers moving through the network.  These walkers can readily adapt to link
weights and link directions, no modifications to the underlying algorithm are necessary.

The second reason for adopting Infomap relates to a result from Park, Lee, and
Gonz\'alez \cite{park2010eigenmode}.  They present the at-first-glance
paradoxical result that a random walk model on an empirically-derived mobility
network does well at reproducing macroscopic phenomena such as the gyradius.
This seems surprising given that human beings have long-term memory, and
consistently travel between fixed sets of destinations
\cite{gonzalez2008understanding}, unlike a diffusing random walker.  Even in
\cite{song2010limits} it was shown, using estimates of the Kolmogorov entropy of
a human trajectory, that there is more information in a trajectory than
estimated by the Shannon entropy alone.  (The resolution of this paradox is to note
that a random walker exploring a new space of locations will not be able to
\emph{generate} mobility networks such as those we derive from the mobile phone
data; a more complex modeling framework is necessary \cite{song2010modelling}.)
Thus, while there is information in a human trajectory beyond the mobility
network, the mobility network still captures a great deal of mobility phenomena.
Infomap's theoretical basis exactly matches this
random-walker-on-a-mobility-network model.

\subsection{Additional properties of mobility habitats}\label{subsec:propertieshabitats}

\begin{figure}[]
    \centerline{\includegraphics[]{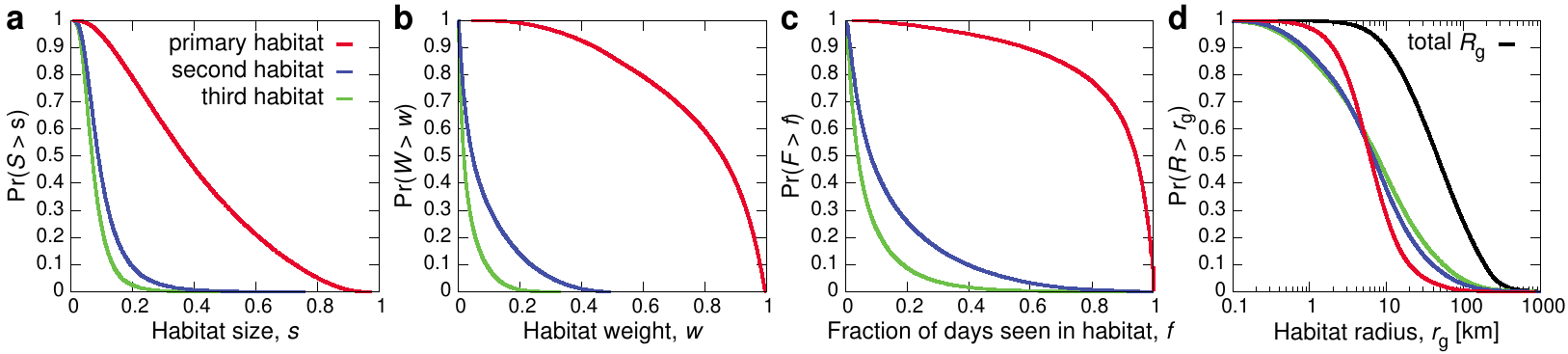}}
    \caption[Additional habitat statistics]{ %
        \textbf{Additional habitat statistics.} %
        We plot the complementary cumulative distributions of habitat size
        (fraction of unique locations within the habitat), habitat weight
        (fraction of calls placed from within the habitat), the fraction of days
        during the nine-month window where the user was active and was observed
        in the habitat, and the habitat radius of gyration. The gyradii
        distributions were also shown in main text Fig.~2a. We see that the
        primary habitat occupies most locations, the majority of phone activity,
        and that most users are found in the habitat nearly every day. Thus the
        primary habitat captures the majority of user dynamics.
        \label{fig:additionalHabitatStatistics}}
\end{figure}

We remark on several additional features of mobility habitats not fully discussed in
the main text.
In Fig.~\ref{fig:additionalHabitatStatistics} we plot the distributions over the
sampled user population of several quantities of interest for the habitats.
These are the ``size'' of each habitat, as given by the number of unique
locations within the habitat; the ``weight'' of each habitat, given by the
fraction of phone calls placed from locations within the habitat; the fraction of
days during the nine-month sample window where the user was active making calls
and observed in the habitat; and the distribution of habitat spatial extent,
given by the gyradius. 
We see that the primary habitat tends to contain most locations and the
overwhelming majority of call activity, and that many users appear in their
primary habitat almost every day.  Thus the primary habitat alone tends to capture the
intrinsic or typical day-to-day activity of a user.

\begin{figure}[t!]
    \centerline{\includegraphics[]{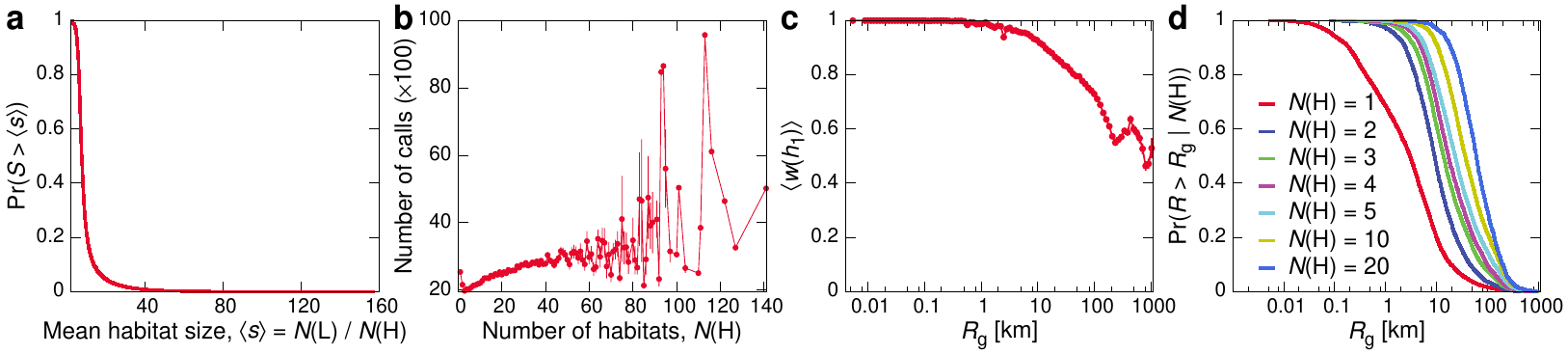}}
    \centerline{\includegraphics[]{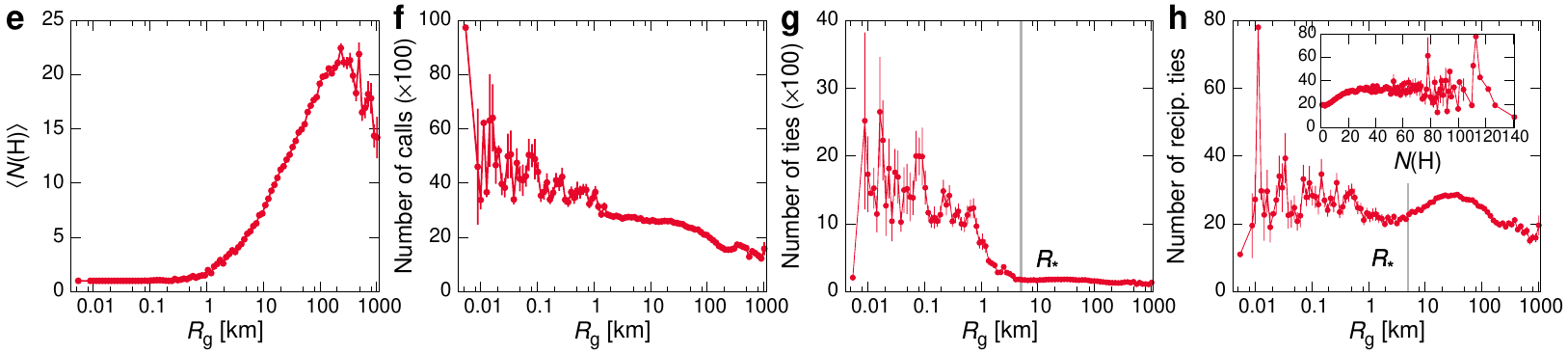}}
    \caption[More habitat properties]{ %
        \textbf{More habitat properties.} %
        \lett{a} %
        The distribution of the average number of locations per habitat is
        narrowly peaked around 7--8 locations per habitat, with only a handful
        of habitats containing large numbers of locations.
        \lett{b} %
        The total number of calls placed by users as a function of the number of
        habitats. There is a weak increasing trend.
        \lett{c} %
        The fraction of call activity $w(h_1)$ occurring in the primary habitat
        as a function of $\Rg$. The majority of activity occurs within the
        primary habitat. This slowly drops as $\Rg$ grows, though the trend is
        weak.
        \lett{d} %
        The distribution of $\Rg$ for users with a given number of habitats
        $N(H)$. We see that users with a single habitat tend to have smaller
        total $\Rg$ than users with multiple habitats, whereas the distributions
        of $\Rg$ change less for users with multiple habitats as the number of
        habitats increases.
        \lett{e} %
        The average number of habitats vs.~$\Rg$. Compare this with main text
        Fig.~2B.
        \lett{f} %
        The average total number of calls shows a slight, relatively constant
        downward trend with $\Rg$.
        \lett{g} %
        The average number of ties vs.~$\Rg$. We see that users with small $\Rg$
        show a sharp increase in the number ties, an effect stronger than the
        increase in the number of calls shown in f.
        \lett{h} %
        The average number of reciprocated ties vs.~$\Rg$. We see no consistent
        trend as $\Rg$ varies.  This means that the increase in the fraction of
        reciprocated ties for users with $\Rg > \Rs$, shown in main text
        Fig.~5D, is due to the decrease in the number of non-reciprocated ties
        as $\Rg$ increases, show here in g.
        (Inset) The number of reciprocated ties as a function of the number of
        habitats $N(H)$. We see a dependence for $N(H) < 20$. 
        Error bars represent $\pm 1$ s.e.
        \label{fig:moreHabitatStatistics}}
\end{figure}

We explore more relationships between habitats, spatial features, and call
activity in Fig.~\ref{fig:moreHabitatStatistics}. We begin in
Fig.~\ref{fig:moreHabitatStatistics}a by plotting the distribution of the
average number of locations per habitat. This distribution is very sharply
peaked around 7--8 locations per habitat, implying a typical habitat size (in
numbers of locations, not spatial extent).  In
Fig.~\ref{fig:moreHabitatStatistics}b we consider the average number of calls
(either voice or text) placed by users as a function of their total number of
habitats.  We see a slow increasing trend. 

Meanwhile, a number of user features can be related to the total gyradius $\Rg$.
In Fig.~\ref{fig:moreHabitatStatistics}c we study habitat weight vs.~$\Rg$. We
see that this weight is approximately 1 until $\Rg \approx \Rs$, after which it
slowly decays to around 0.5, on average. This means that users tend to be less
concentrated within their main habitat as their movements grow, which makes
sense.
We return to the distribution of $\Rg$ in Fig.~\ref{fig:moreHabitatStatistics}d
but now consider it conditioned on users having a specific number of habitats
total.  We see that users with a single habitat tend to have far smaller $\Rg$
than users with more than one habitat, which is reasonable, while users with
three habitats have a similar $\Rg$ distribution as users with only two
habitats.
A related quantity is the average number of habitats as a function of $\Rg$,
shown in Fig.~\ref{fig:moreHabitatStatistics}e. This number is essentially flat
at one habitat until $\Rg \approx 1$ km, after which it slowly grows with $\Rg$.

It is especially interesting to compare social activity with $\Rg$, as was done
in main text Fig.~5. Doing so further sheds light on some of the main text
results regarding $\pmfc$ and $f_\mathrm{reciprocal}$, the fraction of ties that
are reciprocated.
First a question of data sparsity. It seems reasonable that users with very low
$\Rg$ may simply not use their mobile phones much. In
Fig.~\ref{fig:moreHabitatStatistics}f we show that this is not the case by
computing the average number of calls as a function of $\Rg$: users with low
$\Rg$ actually tend to make more calls and this call volume slowly decays as
$\Rg$ grows. Perhaps this means that low $\Rg$ users have more social contacts
than users with higher $\Rg$? To see this we plot in
Fig.~\ref{fig:moreHabitatStatistics}g the number of unique call recipients as a
function of $\Rg$ and we see a rather dramatic increase in these ties at or slightly
before $\Rg \approx \Rs$. This means that users tend to have much more diverse
calling patterns when they seldom move, although we caution that the statistics
are not plentiful at the extreme range of $\Rg$. 
Likewise, an important fact is that these ties may not all be meaningful. As
mentioned in the Materials and Methods section of the main text, we estimate a
tie to be meaningful if it is reciprocated. In
Fig.~\ref{fig:moreHabitatStatistics}h we see that the average number of
reciprocated ties is approximately independent of $\Rg$, implying that, while
many different activity features are affected by mobility phenomena, the number
of meaningful social ties may be constant regardless. (On the other hand, we do
see a weak dependence on the number of habitats, see inset.) 
The famous Dunbar's number~\cite{dunbar1992neocortex} fixing the maximum size of
one's social circle is a relevant quantity of interest.

\begin{figure}[t]%
    \begin{minipage}[c]{2cm}\centering%
        {\includegraphics[trim= 0 0 120 0,clip=true]{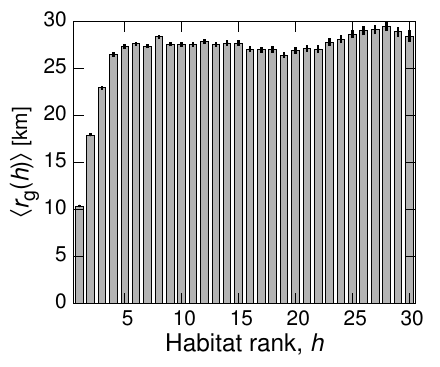}}
    \end{minipage}\hfill
    \begin{minipage}[c]{11.5cm}%
        \centering%
        \caption[Habitat gyradius versus habitat rank]{%
            \textbf{Habitat gyradius versus habitat rank.}
            We see that the most active habitats tend to be more compact spatially,
            on average, than those less frequented habitats.  The overall habitat
            size quickly saturates at around $25-30$km, however, indicating an 
            intrinsic maximum spatial extent.  Error bars represent $\pm 1$ s.e.
            \label{fig:rgVsh}}%
    \end{minipage}%
\end{figure}

Another spatial feature we study here is the spatial extent of habitat $h$, captured by its
gyradius $\rg(h)$, as a function of the rank $h$ of the habitat.  We show in
Fig.~\ref{fig:rgVsh} that more active (lower $h$) habitats tend to be smaller on
average, going from $\rg\approx 10$ km to $\rg \approx 20$ km as $h$ goes from
$1$ to $3$.  However, we see a rapid saturation in $\rg$ to values typically
between $25$ to $30$ km, for $h>5$.  This indicates that there is an
\textbf{intrinsic upper bound} on effective habitat spatial extent, further
emphasizing their cohesive nature.  
              
\begin{figure}[t]
    \centerline{\includegraphics[]{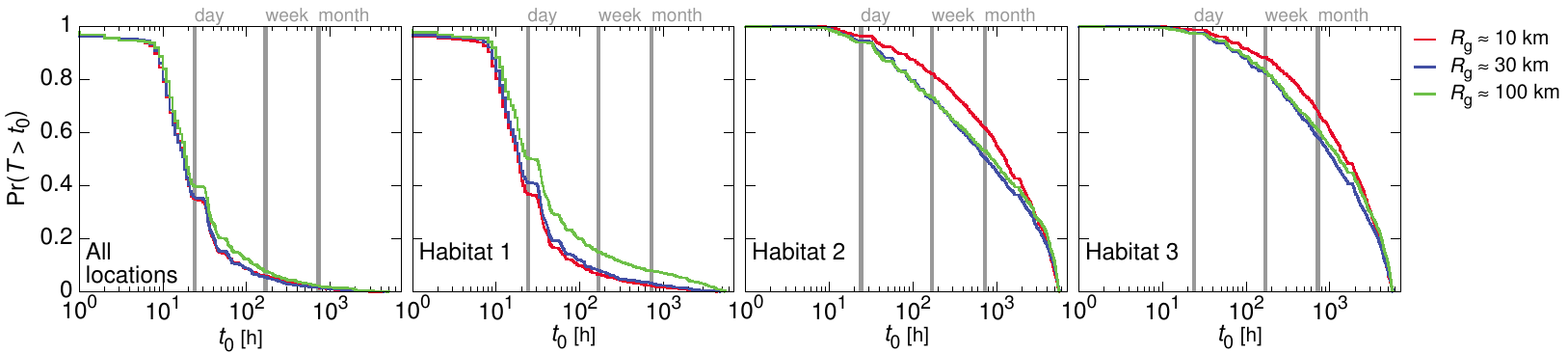}}
    \caption[Population distributions of habitat arrival times]{%
        \textbf{Population distributions of habitat arrival times.} 
        The distributions of times $t_0$ when each user placed his or her first
        call from any location or from within a habitat.  Different
        curves represent groups of users with different total gyradius $\Rg$.
        We see that most users place their first call within Habitat 1 very
        early, often within a day of the start of data collection. As $\Rg$
        grows, these distributions change only slightly, with far ranging
        travelers having a slightly higher probability of delaying their first
        appearance in Habitat 1 (green curve). 
        Habitats 2 and 3 show longer waits until users arrive at these
        locations.  Interestingly, there is only a minor dependence on $\Rg$:
        users with smaller values of $\Rg$ (red curve) tend to wait slightly
        longer to enter these habitats.
        \label{fig:distrHabitatArrivalTimes}}
\end{figure}

Finally, we also study the distribution of habitat entrance times
$t_0(h)$, the time it takes for the user to first enter a location within
habitat $h$ ($t=0$ is the time of the user's first call). In the main text we
show how the delay in entering habitats greatly alters the temporal scaling in
$\rg$, so that habitats grow only logarithmically in time, distinct from the
polylogarithmic growth reported in the literature for the full mobility
pattern~\cite{gonzalez2008understanding,song2010modelling}.  We present in
Fig.~\ref{fig:distrHabitatArrivalTimes} the complementary cumulative
distributions of $t_0$ for the first three habitats, as well as the time it
takes for the first call to occur (which can be thought of as $t_0$ for all
locations).  We see that the distribution for Habitat 1 is functionally similar
to the total distribution, while $t_0$ for Habitats 2 and 3 tends to be higher
in value.  We see a minor dependence on the total mobility extent, quantified
with $\Rg$: Users with higher $\Rg$ tend to wait longer before entering Habitat
1, perhaps because they are more likely to be traveling far from home when data
collection begins, while users with lower $\Rg$ tend to wait slightly longer
before entering Habitats 2 or 3, perhaps because they tend to travel less
frequently.  These results further emphasize the fundamental role that mobility
habitats play in determining the magnitude and dynamics of human mobility and
travel patterns.

\section{Demographic and communication effects} \label{sec:demogs}

\begin{figure}
    \centering 
    \includegraphics[]{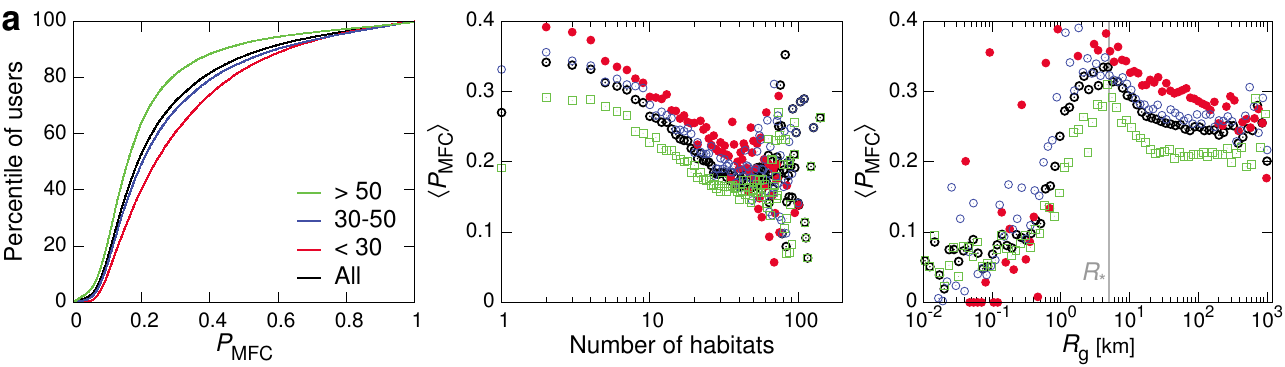} 
    \includegraphics[]{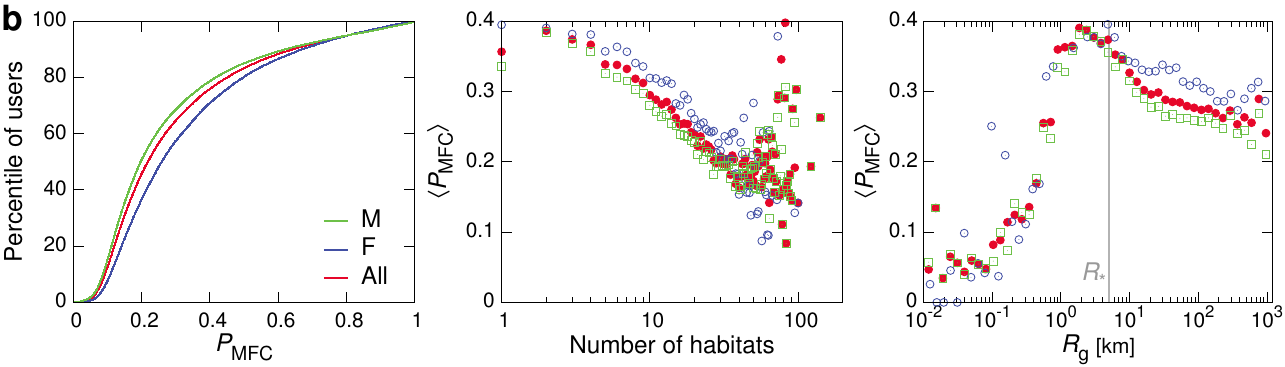}  
    \includegraphics[]{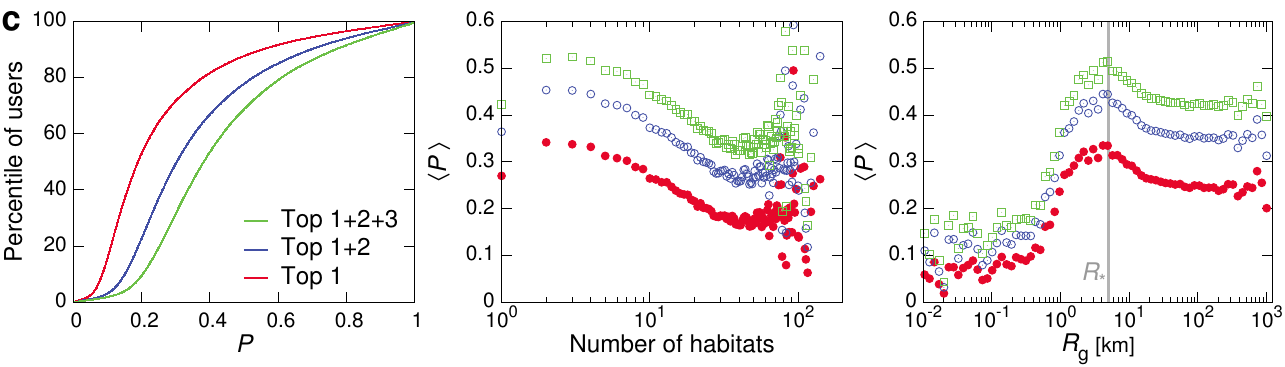}
    \caption[Demographics and extensions of interaction concentration]{%
        \footnotesize
        \textbf{Demographics and extensions of interaction concentration.}
        \lett{a} %
        Age groups. Users older than fifty tend to be less concentrated on their
        MFC, while younger generations focus more on their MFC. (``All''
        indicates all users with self-reported age.)
        \lett{b} %
        Gender groups. Female users call slightly more to their MFC than male
        users. (``All'' indicates all users with self-reported gender.)
        \lett{c} %
        The cumulative probabilities for calling the top-two and top-three most
        frequently contacted friends exhibit similar trends as $\pmfc$ with
        respect to the mobility measures.
        Overall, the primary difference for the demographic groups is the
        average value of their respective $\pmfc$.  After accounting for this,
        we observe the same relationships between $\pmfc$ and mobility.
       \label{fig:age_sex_cmfc}}
\end{figure}

\begin{figure}[t!]
    \centering 
    \includegraphics[]{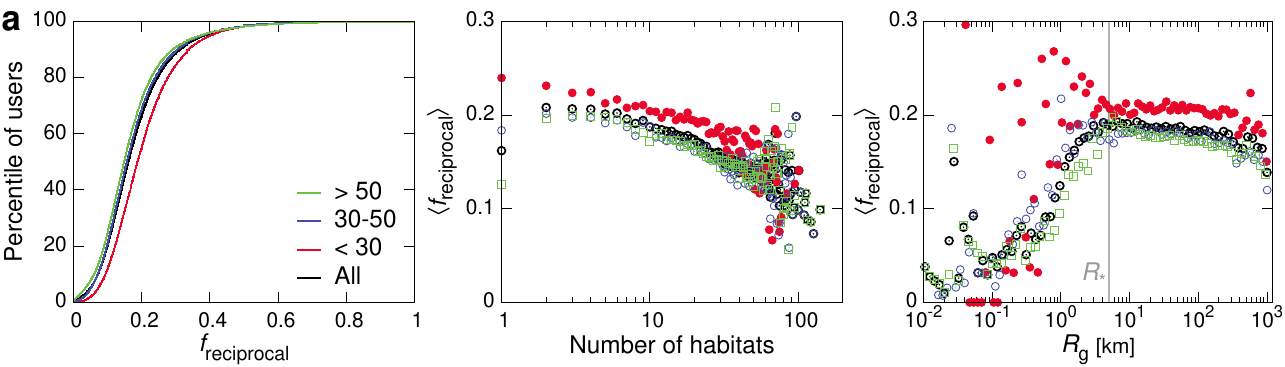} 
    \includegraphics[]{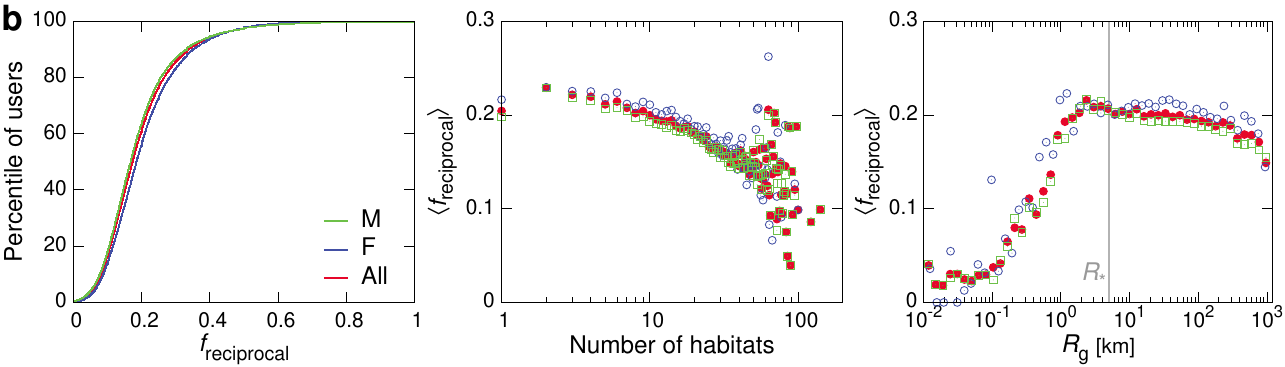}  
    \caption[Demographics and tie reciprocity]{
        \footnotesize
        \textbf{Demographics and tie reciprocity.}
        \lett{a} %
        Age groups. Users under the age of 30 tend to have slightly more
        reciprocated ties than older users. 
        \lett{b} %
        Gender groups. The fraction of reciprocated ties shows no dependence on
        user gender.
    \label{fig:frecip_age_sex}
    }
\end{figure}

We decompose the sample population by self-reported age and gender, and present
the results from main text Fig.~5 with respect to different age and gender
groups. Results for the different groups are shown in
Fig.~\ref{fig:age_sex_cmfc}a and \ref{fig:age_sex_cmfc}b. We see the same
overall features for these groups, with some small quantitative differences,
that we observed in the main text for the entire population.
We also compare in Fig.~\ref{fig:age_sex_cmfc}c the probability of calling the
most frequent contact ($\pmfc$) with the cumulative probabilities of calling the
top-two and top-three most frequently contacted communication partners. We see
that the cumulative probabilities exhibit a similar trend as $\pmfc$ with
respect to mobility, suggesting that the relationship between mobility patterns
and interaction concentration captured by $\pmfc$ is stable over the most
frequently contacted partners. 
Finally, in Fig.~\ref{fig:frecip_age_sex} we study the fraction of ties that are
reciprocated $\frecip$ for different age and gender groups. We see a small
increase in $\frecip$ for users under the age of 30, compared with older users,
while there is no dependence on gender.

\section{Data sparsity} \label{sec:sparsity}

There is an important factor to consider when using mobile phone data and that
is how phone usage affects measurements, as data are available only when users
engage their mobile phones.  While we select active users using the criteria
of~\cite{song2010limits}, specifically intended to mitigate such problems, there
still exist many time periods where a user does not use the phone and thus we do
not have any information.  We now study this in further detail.

\begin{figure}%
\begin{minipage}[c]{9cm}\centering%
    {\includegraphics[trim=0 0 0 15, clip=true]{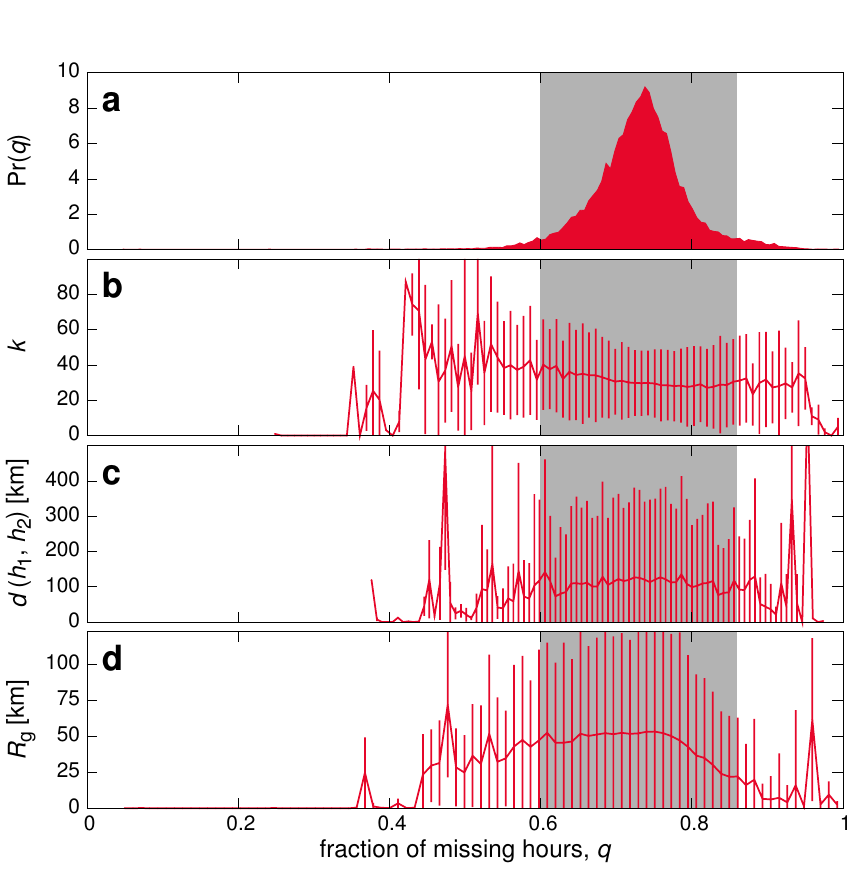}}
\end{minipage}\hfill
\begin{minipage}[c]{7.3cm}%
	\centering%
    \caption[Missing data do not affect most statistics]{%
        \textbf{Missing data do not affect most statistics}.  
        Mobile phone data is inherently incomplete since data is only available
        when users place calls.  We apply the selection criteria used
        in~\cite{song2010limits} to mitigate the missing data issue.  When
        considering the fraction of hours $q$ without data out of the 34 weeks,
        we see that users are typically missing data $\approx 75\%$ of the time
        (\textbf{a}).  One may expect this to cause bias, yet for a number of
        measures (\textbf{b}--\textbf{d}) we see almost no trend in the shaded
        region (containing $\approx 95\%$ of the population).  Only for the
        gyradius (\textbf{d}) do we see a minor dependence for higher values of
        $q$.  These results indicate that our measures are not sensitive to
        missing data. Error bars indicate $\pm 1$ s.d.
    \label{fig:missingdataTest}}
\end{minipage}
\end{figure}

We compute for each user the fraction of hours $q$, out of the nine-month
window, where the user is not active.  In Fig.~\ref{fig:missingdataTest} we plot
the distribution of $q$ over the $90$k users; we see that users are inactive on
average around $75\%$ of the time.  (This distribution, and its consequences,
was also discussed in \cite{song2010limits}.)
While this may seem problematic at first, we are able to proceed because we are
integrating over such a long time window, extracting robust amounts of data for
the quantities we are interested in.  For example, in
Fig.~\ref{fig:missingdataTest} we also plot the number of communication partners $k$, the
distance between the first and second habitats $d(h_1,h_2)$, and the total
gyradius $\Rg$, all as a function of the missing fraction $q$. We see almost
no trend or dependence on $q$.  Meaning that users missing data $40\%$ of the
time give the same or similar statistics as users missing data $80\%$ of the
time, for example.  Only for the gyradius do we see a small drop in
$\Rg$ for larger values of $q$.  We note, however, that even this trend remains
within 1 s.d.~and is thus not significant.

\section{Controls and hypothesis tests} \label{sec:controls}

We introduce Habitat controls to determine how meaningful the discovered
habitats are, where we use different groupings of mobility locations to form
control or null habitats.  To compute habitat controls we form randomized
habitats by shuffling locations between the original habitats at random in such
a way as to preserve for each user both the number of habitats and the number of
locations per habitat, strictly controlling for the size distributions of the
habitats.  These control for the nature of the groupings we find and whether or
not it is meaningful for particular locations to share a habitat.  

\begin{figure}
    \centering
        \includegraphics[]{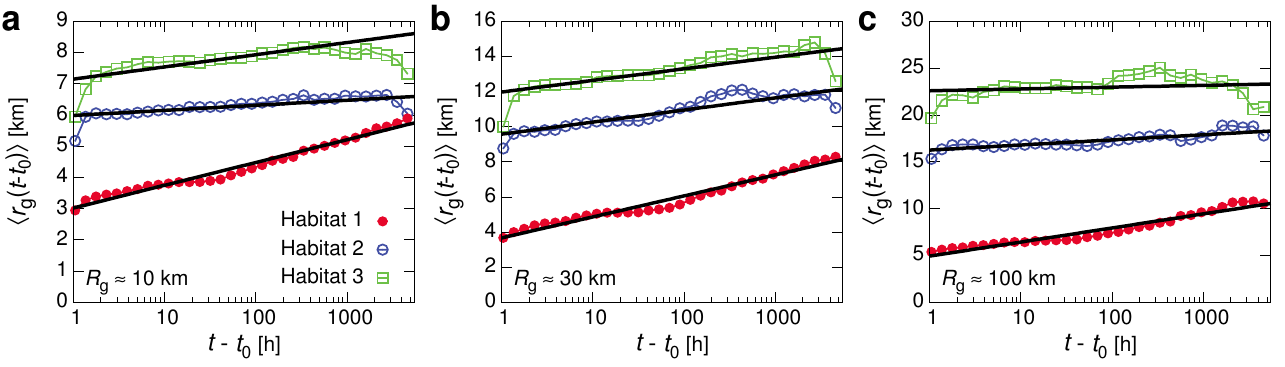}
        \includegraphics[]{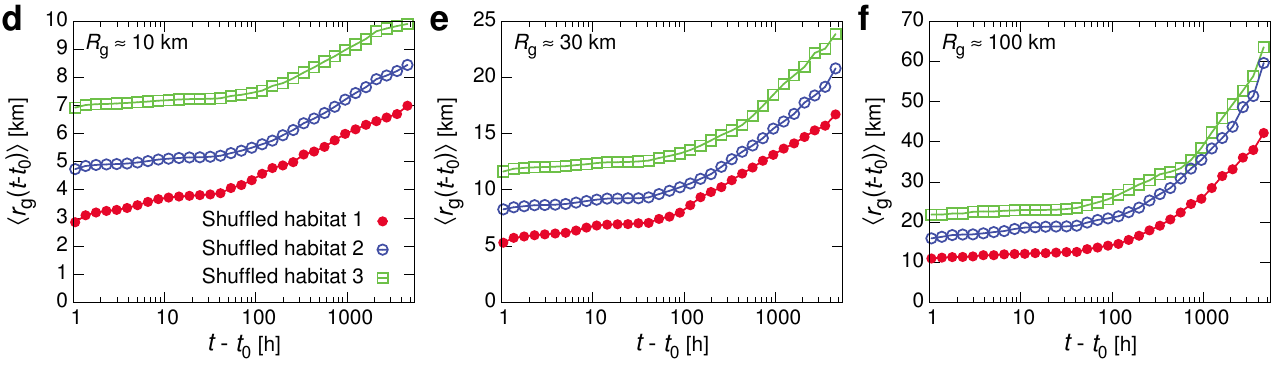}
    \caption[Temporal evolution of gyradius for real and null habitats]{%
        \textbf{Temporal evolution of gyradius for real and null habitats.}
        \textbf{(a-c)} Real habitats.
        The temporal evolution of $\left<\rg(h)\right> \sim \log
        \left(t-t_0\right)$ in all cases for user groups with total $\Rg
        \approx$ $10$, $30$, and $100$ km. Straight lines indicate fits of the form
        $\rg = A\log \left(t-t_0\right) + B$, for constants $A$ and $B$.
        \textbf{(d-f)} Null habitats.
        The same as a-c but for randomized control habitats, constructed for
        each user by randomly reshuffling locations between habitats while
        preserving the number of habitats and the number of locations within
        each habitat.  We see the pure logarithmic scaling of $\rg$ is lost.
        \label{fig:temporal_rg_others_shuffled}}
\end{figure}

To begin, in Fig.~\ref{fig:temporal_rg_others_shuffled}a-c we show the temporal
evolution of the gyradius $\rg(t)$ vs.~$t$ for the first three habitats.  We
study three sets of users, with total $\Rg \approx 10$, $30$, and $100$ km,
respectively.  For all groups we see that $\rg \sim \log(t-t_0)$, amplifying the
results from main text Fig.~3d.  Meanwhile, in
Fig.~\ref{fig:temporal_rg_others_shuffled}d-f we show the same quantities but
for the shuffled habitats where locations are randomized.  We see that the pure
logarithmic time evolution is lost, indicating that the evolution we observe is
not due to the relative sizes (numbers of locations) of the habitats, nor  to
simply the number of habitats, but due more fundamentally to their spatial
structure and the spatiotemporal flows of the users.                    

Meanwhile, the results from main text Fig.~5 show an intriguing relationship
between human mobility and communication activity. Here we quantify the relationship
using the Kendall's $\tau$ (tau-b) rank correlation coefficient \cite{kendall90rank},
which is a nonparametric hypothesis test used to measure the association between
two measured quantities.  The coefficient $\tau>0$ indicates positive
association, $\tau<0$ indicates negative association, and $\tau=0$ indicates the
absence of association.

Mobility is evaluated by the number of habitats $\Nh$ and the total gyradius
$\Rg$, while interaction concentration is quantified by the probability of calling
the most frequent contact $\pmfc$, the cumulative probability of calling the
top-three most frequently contacted partners $\cmfc$, and the total number of
partners $\Nf$.

In Table~\ref{tab:corr_mob_soc}, we see a negative association between $\Nh$ and
$\pmfc$ (as well as $\cmfc$), while $\Nh$ and $\Nf$ are positively correlated.
This suggests that people who are more habitually mobile (with more habitats)
tend to distribute their communication over more contacted ties. 

An association between $\Rg$ and the communication measures at first appears absent.
However, when separating users who possess only a single habitat ($\Nh=1$) from
those who don't ($\Nh>1$), we discover that the relationship with $\Rg$ shows
two opposing trends: for users with a single habitat, $\Rg$ grows with $\pmfc$;
when users have more than one habitat, their $\pmfc$ begin to drop. These
correlations suggest a coupling between mobility habitats and interaction
concentration, which cannot be captured by a single $\Rg$ value. Similarly, for
$\frecip$ the association with $\Rg$ is stronger when $\Nh = 1$, than when $\Nh
> 1$.

\begin{table}
    \centering
    \caption[Nonparametric correlations between mobility and communication]{%
        \textbf{Nonparametric correlations between mobility and communication.}
        We quantify the relationship between mobility (columns) and
        communication (rows) using
        Kendall's $\tau$ rank correlation coefficient.  User mobility is
        evaluated by the number of habitats $\Nh$ and the gyradius $\Rg$, and
        their concentration is evaluated by $\pmfc$, $\cmfc$ (the cumulative
        probability of calling any of the three most frequently contacted
        partners) and $\Nf$ (the number of partners). Each table entry shows the
        coefficient $\tau$ and its corresponding $p$-value (parenthesis). We see
        a negative association between $\Nh$ and $\pmfc$ (and $\cmfc$), while
        $\Nh$ and $\Nf$ are positively correlated. An association between $\Rg$
        and the communication measures appears absent. However, when separating users
        with a single habitat ($\Nh=1$) from the rest ($\Nh>1$), we see that the
        growth of $\Rg$ has two opposite trends: within a single habitat, $\Rg$
        grows with $\pmfc$; when users have more than one habitat, $\pmfc$ begin
        to drop.  This implies a coupling between mobility habitats and
        interaction concentration, one that cannot be captured by $\Rg$ alone.
    \label{tab:corr_mob_soc}}
{\footnotesize
\begin{tabular}{lllll}
    \toprule
& \multicolumn{4}{c}{Mobility}\\
 \cmidrule(r){2-5}
{Communication} & $\Nh$     & $\Rg$     & $\Rg~(\Nh=1)$ & $\Rg~(\Nh>1)$          \\
    \midrule                                             
$\pmfc$&    $-0.133~(<2.2\e{-16})$ & $-0.00648~(0.0039) $ &$0.354~(<2.2\e{-16})$ & $-0.0324~(<2.2\e{-16})$\\
$\cmfc$&    $-0.153~(<2.2\e{-16})$ & $-0.0127~(1.7\e{-8}) $ &$0.358~(<2.2\e{-16})$ & $-0.0369~(<2.2\e{-16})$\\
$\Nf$    &   $0.214~(<2.2\e{-16})$ & $-0.0728~(<2.2\e{-16}) $  &$-0.44~(<2.2\e{-16})$ & $-0.0619~(<2.2\e{-16})$\\
$\frecip$& $-0.0452~(<2.2\e{-16})$ & $ 0.034~(<2.2\e{-16}) $  &$0.369 ~(<2.2\e{-16})$ & $0.00462~(0.050929)$\\
   \bottomrule
\end{tabular}}
\end{table}

{\small
\addcontentsline{toc}{section}{References} 

}